\documentclass[aps,twocolumn,showpacs,preprintnumbers,amsmath,amssymb,pra]{revtex4}

\usepackage{graphicx}
\usepackage{epsfig}
\usepackage{amsmath}
\usepackage{amsfonts}
\usepackage{mathrsfs}
\begin{document}

\title{Universal local pair correlations of Lieb-Liniger bosons at quantum criticality}

\author{M.-S. Wang$^{1,2}$}
\author{J.-H. Huang$^{1}$}
\author{C.-H. Lee$^{1}$}

\author{X.-G. Yin$^{3}$}
\author{X.-W. Guan$^{4,5}$}
\author{M. T. Batchelor$^{5,6}$}
\affiliation{$^{1}$State Key Laboratory of Optoelectronic Materials and Technologies, School of Physics and Engineering, Sun Yat-Sen University, Guangzhou 510275, China}

\affiliation{$^{2}$Department of Physics, Ludong University, Yantai 266000, China}

\affiliation{$^{3}$Division of Materials Science, Nanyang Technological University, Singapore 639798}

\affiliation{$^{4}$Wuhan Institute of Physics and Mathematics, Chinese Academy of Sciences, Wuhan 430071, China}

\affiliation{$^{5}$Department of Theoretical Physics, Research School of Physics and Engineering, Australian National University, Canberra ACT 0200, Australia}

\affiliation{$^{6}$Mathematical Sciences Institute, Australian National University, Canberra ACT 0200, Australia}

\begin{abstract}

The one-dimensional Lieb-Liniger Bose gas is a prototypical many-body system featuring universal Tomonaga-Luttinger liquid (TLL) physics
and free fermion quantum criticality. We analytically calculate finite temperature local pair correlations for the strong coupling 
Bose gas  at quantum criticality using the polylog function in the framework of the Yang-Yang  thermodynamic equations. 
We show that the local pair correlation has the universal value $g^{(2)}(0)\approx 2 p/(n\varepsilon)$ in the quantum critical regime, 
the TLL phase and the quasi-classical  region, where $p$ is the pressure per unit length rescaled by the interaction energy 
$\varepsilon=\frac{\hbar^2}{2m} c^2$ with interaction strength $c$ and linear density $n$. 
This suggests the possibility to test finite temperature local pair correlations for the TLL in  the relativistic dispersion regime and to 
probe quantum criticality with the local correlations beyond the TLL phase. Furthermore, thermodynamic properties at high 
temperatures are obtained by both high temperature and virial expansion of the Yang-Yang  thermodynamic equation.

\end{abstract}

\volumeyear{year}
\volumenumber{number}
\maketitle

\section{Introduction}

Advances in the trapping and cooling  of  atoms in optical waveguides have opened up exciting possibilities for testing theory in
low-dimensional quantum systems. Observed results to date are seen to be in excellent agreement with results obtained
using the mathematical  methods and analysis of exactly solved models \cite{Lieb,Gaudin,Yang}.
These include the remarkable experimental realization of the Tonks-Girardeau gas \cite{Paredes,Kinoshita,Exp2},
the super Tonks-Girardeau gas \cite{Haller},  Yang-Yang thermodynamics on an atom chip \cite{Amerongen},
and the phase diagram  of the attractive Fermi gas \cite{Liao}. Such exquisite  tunability with tightly confined ultracold
atoms provides  unprecedented opportunities  for improving our understanding of
novel quantum  phenomena such as  quantum criticality, universal scaling theory, spin-charge separation and
Tomonaga-Luttinger liquid (TLL) physics.

The one-dimensional (1D) delta-function interacting Lieb-Liniger
Bose gas \cite{Lieb} is a many-body system solved exactly by the
Bethe ansatz hypothesis. It has had a tremendous impact as an
archetypical system in quantum statistical mechanics
\cite{Cazalilla}. It's $R$-matrix \cite{McGuire} provides the
simplest solution of the Yang-Baxter equation \cite{Jimbo}. Yang
and Yang \cite{Yang-Yang} showed that  the  thermodynamics of
Lieb-Liniger  bosons can be determined from the minimisation of
the Gibbs free energy subject to the Bethe ansatz equations
\cite{Korepin,Takahashi-b}. This thermodynamic Bethe ansatz method
has been extended to a wide range of 1D  quantum many-body systems
\cite{Takahashi-b}. In particular, Yang-Yang thermodynamics is
fundamental to the $Y$-system which has emerged as a ubiquitous
integrable structure in mathematical physics \cite{Kuniba}. In the
present context, it provides the framework to study
thermodynamics, quantum criticality and TLL  physics. The
equation of state for Lieb-Linger bosons has been obtained
\cite{GB} analytically for strong coupling and low temperature in
terms of the polylog function. The expression for the equation of
state in this regime enables the exploration of TLL physics and
quantum criticality in this system.

In a recent experiment \cite{Armijo}, thermal fluctuations were studied in a highly elongated  weakly interacting Bose gas  at  high temperatures,
where the quantum fluctuations are strongly suppressed. In this regime, the measured thermal fluctuations are in good agreement with the exact
Yang-Yang thermodynamics.  However, at quantum criticality, where the temperature is very low and interaction is very strong,  quantum fluctuations are strongly enhanced. 
Towards the quantum critical regime, phonon fluctuations have been observed in the regime where the temperature is less than the chemical potential \cite{Armijo2}. 
It is of particular interest to understand
quantum correlations and fluctuations at quantum criticality. A finite temperature quantum phase transition does not exist in the
1D Lieb-Liniger Bose gas. However,  there is a critical point in the grand canonical ensemble, when the chemical potential $\mu_c=0$,
which separates the vacuum from a filled ``Fermi sea"  of particles at zero temperature.  At finite temperatures, a TLL with relativistic dispersions can be sustained
in a region of the $T-\mu$ plane.  This implies that for temperatures below a crossover value $T^*$, the low-lying excitations have a linear relativistic dispersion relation.
If the temperature exceeds this crossover value, the excitations involve free quasiparticles with non-relativistic dispersion. This crossover temperature can be identified
from the breakdown of linear temperature-dependent entropy, see Fig.~1. In this phase diagram, quantum criticality is in the regime where
$t=k_BT/(\frac{\hbar^2}{2m}c^2)$ is small, but   $k_BT>|\mu-\mu_c|$, or for the temperature  below the degenerate temperature $k_BT<\frac{\hbar^2}{2m}n^2$.
Here $k_B$ is the Boltzmann constant.

\begin{figure}[tbp]
\includegraphics[width=1.150\linewidth]{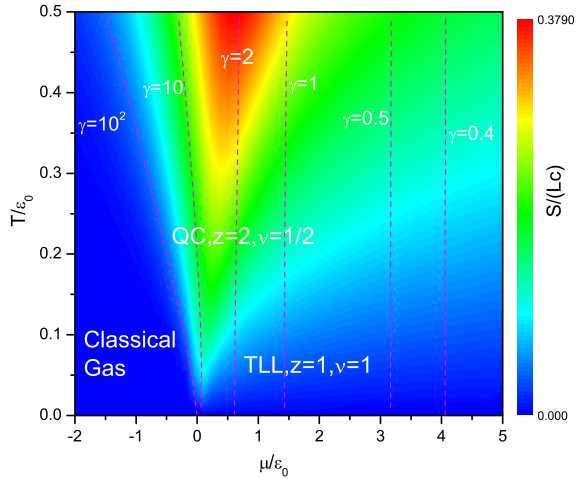}
  \caption{Quantum phase diagram of Lieb-Liniger bosons. The plot shows the dimensionless  entropy in the $T-\mu$ plane
  obtained from the Yang-Yang thermodynamics (\ref{TBA}), with $\varepsilon_0=\frac{\hbar^2}{2m}c^2$. 
  A crossover temperature separates  the TLL critical phase with
  dynamic exponent $z=1$ and correlation length exponent $\nu=1$ from the quantum critical regime (QC) 
  with exponents $z=2$ and $\nu=1/2$ near the critical point $\mu_c=0$.
  Another crossover temperature separates the quasi-classical gas from the quantum critical regime. 
  The red dashed lines indicate the values of the dimensionless coupling strength $\gamma$. }
\label{fig:scaling}
\end{figure}

The local two-body (pair) correlations have been analytically calculated for a few limiting cases at $T=0$ \cite{Gangardt}
and at finite temperatures for the TLL phase from the Yang-Yang thermodynamic equations \cite{Kheruntsyan,Cazalilla2,GBT,Mussardo,Kormos}.
However, the local pair correlations have not been derived in the quantum critical regime. Moreover, it is very interesting to find an intrinsic relation  between the local pair correlation and the equation of state.  In principle, the Yang-Yang thermodynamic equations enable the numerical calculation of  the local pair
correlations in the whole parameter space. The local pair correlations are related to the universal contact which measures
the probability of two bosons (or two fermions  with opposite spins) at the same position \cite{Tan,Tan1D,Vignolo2013}.

In the present paper,  we calculate the local pair correlations of
Lieb-Liniger bosons in analytic fashion from the Yang-Yang  thermodynamic equations using the polylog function in order
to understand universal features of quantum fluctuations and critical phenomena in an archetypical quantum system.  The local pair correlation  is  discussed  in the $T-\mu$ phase diagram in terms of quantum criticality.  In addition, a high temperature expansion of the Yang-Yang  thermodynamic equations is obtained in the strong coupling regime.
This allows one to extract the contributions from  quantum and thermal fluctuations to the classical Boltzmann gas at high temperatures.

\section{Model and equation of state}

The Hamiltonian
\begin{equation}
{\cal H}=-\frac{\hbar ^2}{2m}\sum_{i = 1}^{N}\frac{\partial^2}{\partial x_i^2}+\,g_{\rm 1D} \sum_{1\leq i<j\leq N} \delta (x_i-x_j)
\label{Ham-1}
\end{equation}
of the Lieb-Liniger  Bose gas  \cite{Lieb} describes $N$ spinless bosons with mass $m$ constrained by periodic boundary conditions
on a line of length $L$. The contact interactions are governed by the effective 1D coupling constant $g_{1D}=-2\hbar ^{2}/(ma_{1D})$ where
$a_{1D}=\left( -a_{\perp }^{2}/2a_{s}\right) \left[ 1-C\left(a_{s}/a_{\perp }\right) \right] $ is the 1D  scattering length with
$a_{\perp }=\sqrt{2\hbar /m\omega _{\perp }}$ and the numerical constant $C\approx 1.4603$ \cite{Olshanii}.
The dimensionless interaction strength is defined by $\gamma =c/n$ with $c=-2/a_{1D}$, where $n=N/L$ is the linear density.
The interaction strength can be controlled by  tuning either $\omega_{\perp}$ or $a_{s}$ in experiments.

In describing the thermodynamics of the model the key quantity is the dressed energy
\begin{equation}
\epsilon(k) = T\ln(\rho^h(k)/\rho(k))
\end{equation}
which plays the role of excitation energy measured from the energy level $\epsilon(k_{\rm F})=0$, where $k_{\rm
F}$ is the Fermi-like momentum.
The thermodynamics of the model in equilibrium follows from the Yang-Yang  equation \cite{Yang-Yang}
\begin{equation}
\epsilon (k)=
\epsilon^0(k) -\mu-
T\int_{-\infty}^{\infty} dq \, a_2(k-q) \ln(1+{\mathrm e}^{-{\epsilon(q)}/{T}})
 \label{TBA}
\end{equation}
where $\epsilon^0(k)=\frac{\hbar^2}{2m}k^2$ is the bare dispersion,
$\mu$ is the chemical potential and
\begin{equation}
a_2(x)=\frac{1}{2\pi}\frac{2c}{c^2+x^2}.
\end{equation}

The pressure $p(T)$ and the free energy $F(T)$  are given in terms of the dressed energy by
\begin{eqnarray}
p(T)&=&\frac{T}{2\pi}\int_{-\infty}^\infty dk \, \ln(1+\mathrm{e}^{-{\epsilon(k)}/{T}})
\label{Pressure}\\
F(T)&=&\mu n-\frac{T}{2\pi}\int_{-\infty}^\infty dk \, \ln(1+\mathrm{e}^{-{\epsilon(k)}/{T}}).
\label{Free-E}
\end{eqnarray}

The pressure of the strongly coupled gas at finite temperatures has been obtained from  (\ref{TBA})
in the form \cite{GB}
\begin{eqnarray}
p \approx -\sqrt{\frac{m}{2\pi\hbar^2}} T^{\frac{3}{2}} \, \mathrm{Li}_{\frac{3}{2}} (-\mathrm{e}^{{A}/{T}})
 \left[1-\frac{p}{\hbar^2c^3/(2m)}
 \right]
  \label{pressure-polylog}
\end{eqnarray}
where $\mathrm{Li}_s$ is the standard polylog function and
\begin{eqnarray}
A&=&\mu+\frac{2\,p(T)}{c}+\frac{1}{2\sqrt{\pi}c^3}\frac{T^{\frac{5}{2}}}{\left(\frac{\hbar^2}{2m}\right)^{\frac{3}{2}}} \mathrm{Li}_{\frac{5}{2}}
(-\mathrm{e}^{A_0/{T}}).\label{EoS-A}
\end{eqnarray}
Furthermore, the Yang-Yang equation (\ref{TBA}) can be expanded in powers of the dimensionless temperature  $t=k_{\rm B}T/\varepsilon_0$
with $\varepsilon_0=(\frac{\hbar^2}{2m}c^2)$.
Thus from (\ref{pressure-polylog}) the dimensionless pressure
$\tilde{p} = p /(\varepsilon_0 c)$ at finite temperatures follows
as
\begin{eqnarray}
\tilde{p} & \approx&  -\frac{t^{\frac{3}{2}}}{2\sqrt{\pi}}\mathrm{Li}_{\frac{3}{2}} (-\mathrm{e}^{{\tilde{A}}/{t}})
\left[1+\frac{\tilde{T}^{\frac{3}{2}}}{2\sqrt{\pi}}
  \mathrm{Li}_{\frac{3}{2}} (-\mathrm{e}^{{\tilde{A}}/{t}})
  \right]  \label{EoS-E}
\end{eqnarray}
with
\begin{eqnarray}
\tilde{A}&=&\tilde{\mu}-\frac{t^{\frac{3}{2}}}{\sqrt{\pi}}\mathrm{Li}_{\frac{3}{2}} (-\mathrm{e}^{{\tilde{A_0}}/{t}})
+\frac{t^{\frac{5}{2}}}{2\sqrt{\pi}}\mathrm{Li}_{\frac{5}{2}} (-\mathrm{e}^{{\tilde{A_0}}/{t}})
\label{EoS-E-A}
\end{eqnarray}
and
\begin{equation}
\tilde{A}_0=\tilde{\mu}-\frac{t}{\sqrt{\pi}} \mathrm{Li}_{\frac{3}{2}} (-\mathrm{e}^{{\tilde{\mu}}/{t}}).
\end{equation}

The result (\ref{EoS-E}) is essentially a high precision equation of state for Lieb-Liniger bosons at quantum criticality.
We will verify that it  is also valid for the high temperature regime as long as  $k_{\rm B}T \ll  \varepsilon$.
Recalling the phase diagram Fig.~\ref{fig:scaling}, the density and the compressibility can be cast into
universal scaling forms \cite{sachdev,Fisher,Zhou-Ho,Erich,GB}
\begin{eqnarray}
n(T,\mu)-n_0(T,\mu)&\approx&
t^{\frac{d}{z}+1-\frac{1}{\nu
    z}}{\cal{F}}\left(\frac{\mu-\mu_c}{t^{\frac{1}{\nu z}}}\right)
\label{n} \\
\kappa (T,\mu)-\kappa_0 (T,\mu)&\approx&
t^{\frac{d}{z}+1-\frac{2}{\nu z}}{\cal{Q}}\left(\frac{\mu-\mu_c}{t}\right)
\label{kappa}
\end{eqnarray}
near the quantum critical point $\mu_c=0$. Here the dynamic exponent $z=2$ and the correlation
length exponent $\nu=1/2$ with the scaling functions given by
\begin{eqnarray}
{\cal{F}}(x)&=&-\frac{c}{2\sqrt{\pi}}\mathrm{Li}_{\frac{1}{2}} (-\mathrm{e}^x)\\
{\cal{Q}}(x)&=&-\frac{c}{2\varepsilon \sqrt{\pi}}\mathrm{Li}_{-\frac{1}{2}}(-\mathrm{e}^x)
\end{eqnarray}
for $t> |\mu-\mu_c|$ in dimensionless units.
The background density and compressibility in the vacuum are zero, i.e.,   $n_0(t,\mu)=\kappa_0(t,\mu)=0$.
These analytical results provide insight into quantum fluctuations near the quantum critical point.
We will further demonstrate that the finite temperature local pair correlations shed light on quantum critical behaviour.

\begin{figure}[tbp]
\includegraphics[width=1.0\linewidth]{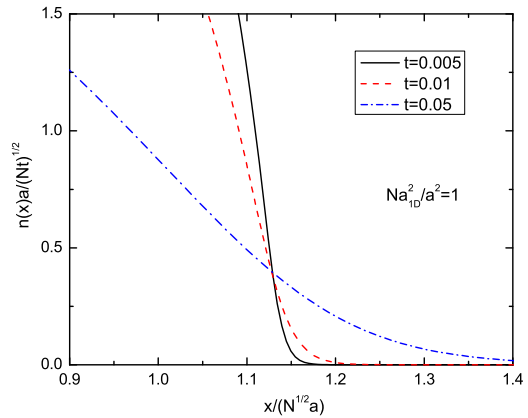}
\caption{The density $n$ vs normalized position at different
temperatures for fixing  $Na_{1D}^{2}/a^{2}=1$. The intersection point  maps out the zero temperature critical point at which
the phase transition from the vacuum to the filled Bose gas occurs in the harmonic trap. Here
$t=T/\protect\varepsilon _{0}$.} \label{densityScaled}
\end{figure}

The thermodynamic Bethe ansatz equations (\ref{TBA}) provide a grand canonical description of the system (\ref{Ham-1}), 
where the chemical potential is fixed.
 Usually, the Lieb-Liniger gas is discussed in the canonical ensemble, i.e., the particle number is fixed.
 In fact, the thermodynamics of a canonical ensemble can be determined from the standard thermodynamic 
 relation $n=\left(\partial p/\partial \mu \right)$, where the pressure per unit length is given by (\ref{Pressure}).
However, in an experiment the quantum gas with fixed number of particles is usually trapped by an
external harmonic potential.
The trapped density varies smoothly along the axial direction, 
with the density distribution read off from the local density function $n(x)$.

Each infinitely small interval around $n(x)$ can be reviewed as a
grand canonical ensemble of the 1D integrable system.  At equilibrium,
the thermodynamics of each interval requests $d \, e(x)  =-pdL+\mu dN$. Through minimization of the total energy of the gas
trapped in the external potential $E=\int e(x)  dx +\int n(x) V(x)$, one can obtain the  thermodynamic condition for equilibrium
\cite{Ma-Yang}
\begin{equation}
\mu +V(x) + \lambda =0
\end{equation}
with $V(x)$ the external trapping potential and $\lambda$ the Lagrange multiplier. 
This equilibrium  condition is nothing but the local density approximation.
 For the Bose gas in  a 1D  harmonic trap, we  thus obtain
 its density distribution from the equation of state within local
density approximation \cite{Imambekov,Menotti}.

Within the local density approximation,
the chemical potentials in the equation of state (\ref{pressure-polylog}) as well as in the TBA
equations (\ref{TBA})  are replaced by the local chemical potentials given by
\begin{eqnarray}
\mu\left( x\right) &=&\mu\left( 0\right) -V\left( x\right). 
\label{LDA1} \label{LDA}
\end{eqnarray}%
Here the external potential is defined as $V\left( x\right)
=m\omega ^{2}x^{2}/2$ with harmonic frequency $\omega $ and the
characteristic length for the harmonic trap is $a=\sqrt{\hbar /m\omega}$.
In this setting, equation (\ref{LDA}) can be alternatively written as
\begin{equation}
\mu \left( y\right) /\varepsilon _{0}=\mu \left( 0\right)/\varepsilon _{0}-y^{2}
\end{equation}%
in which the dimensionless coordinate $y=x/(a^{2}c)$.
 In terms of dimensionless units, the dimensionless density $n/c$ can be
obtain for fixed dimensionless chemical potential $\mu/\varepsilon_{0}$.
The total particle
number $N$ is obtained from the relation
\begin{equation}
\frac{N a_{1D}^{2}}{a^{2}}=4\int_{-\infty }^{\infty }{n\left( y \right) }{c}dy \label{N}
\end{equation}%
with the 1D scattering length $a_{1D}=-2/c$.
For fixed value of trapping centre chemical potential $\mu \left( 0\right) /\varepsilon _{0}$,
  we may determine the value  $N a_{1D}^{2}/{a^{2}}$.
  In turn, for  different  values of  $N a_{1D}^{2}/{a^{2}}$, i.e., for fixed particle number, 
  the thermodynamic properties can be mapped out through the density profiles of the trapped gas at finite temperatures.
  E.g.,  in Fig.~{\ref{densityScaled}}, we show the scaled density
distributions of bosons in the harmonic trap for $ N
a_{1D}^{2}/a^{2}=1$ at different temperatures. It is clearly seen that
the density curves at  different temperatures intersect at a common point.  
We read off the  dynamic exponent $z=2$ and the correlation
length exponent $\nu=1/2$ from the universal scaling function (\ref{n}) within the local density approximation.

\begin{figure}[tbp]
\includegraphics[width=1.050\linewidth]{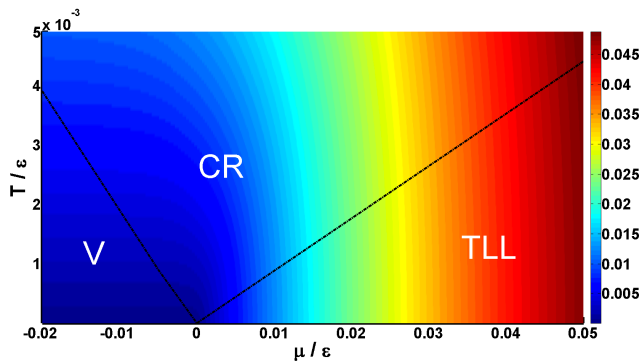}
  \caption{Local pair correlation $g^{(2)}(0)$ at low temperatures in the $t-\mu$ plane.
  The black lines denote the crossover temperature separating the TLL phase from the quantum
  critical regime (CR) and the vacuum (V) from the quantum critical regime (CR). }
\label{fig:g2-all}
\end{figure}

\begin{figure}[tbp]
\includegraphics[width=1.050\linewidth]{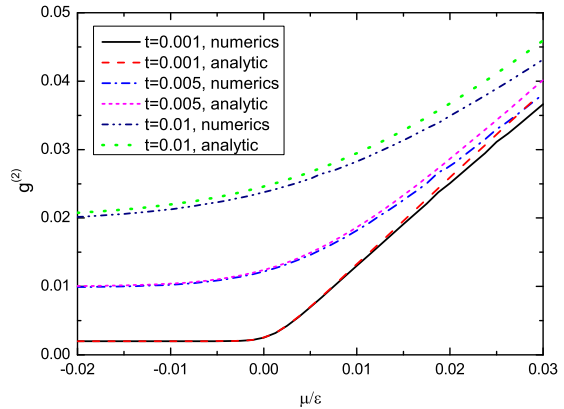}
  \caption{Local pair correlation $g^{(2)}(0)$ vs  chemical potential at quantum criticality.  
  The correlation function  $g^{(2)}(0)$ at  different temperature values shows good  agreement  
  between the analytical relation (\ref{g2-re}) and the numerical result obtained from the TBA equation (\ref{TBA}). }
\label{fig:g2TLL}
\end{figure}

\section{Local pair correlations}

The general calculation of correlation functions in quantum many-body systems is a notoriously difficult problem,
even more so at finite temperature \cite{Korepin}.
Fortunately the two-particle local pair correlation $g^{(2)}(0)$ can be calculated relatively easily from the free energy.
In the grand canonical description, the Hellmann-Feynman theorem gives
\begin{equation}
g^{(2)}(0) := \langle \Psi^{\dagger}\Psi^{\dagger}\Psi\Psi \rangle
=\frac{2m}{\hbar^2n}\left(\frac{\partial
  f}{\partial c }\right)_{n,t}
\end{equation}
where $f=\mu-p/n$ is the free energy per particle.
For constant $n$ the local pair correlations require the calculation of the
derivatives $\partial \mu/\partial c$ and $\partial p/\partial c$, which can be done  by iteration.
Using the result (\ref{pressure-polylog}) we find
\begin{eqnarray}
\frac{\partial p}{\partial c}  \approx -\sqrt{\frac{mT}{2\pi\hbar^2}}  \, \mathrm{Li}_{\frac{3}{2}} (-\mathrm{e}^{{A}/{T}})
 \left[1-\frac{2p}{\hbar^2c^3/(2m)}
 \right]\frac{\partial A}{\partial c}
  \label{pressure-d}
\end{eqnarray}
where  $\partial A/\partial c$ can be calculated from (\ref{EoS-E-A}) by iteration.
The calculation of $\partial \mu/\partial c$ is cumbersome.
To this end, we first find the explicit form of the density $n=\partial p/\partial \mu$
and then take the derivative with respective to $c$. After a lengthy iteration, we find
\begin{widetext}
\begin{eqnarray}
\frac{\partial \mu}{\partial c } &\approx& -\sqrt{\frac{2m}{\pi \hbar^2}}\frac{T^\frac{3}{2}}{c^2}
\left\{ \frac{\mathrm{Li}_{\frac{1}{2}} (-\mathrm{e}^{{A}/{T}})^2}{\mathrm{Li}_{-\frac{1}{2}} (-\mathrm{e}^{{A}/{T}})}
\left[ 1+\frac{1}{c}\sqrt{\frac{2mT}{\pi\hbar^2}}\mathrm{Li}_{\frac{1}{2}} (-\mathrm{e}^{{A}/{T}})
-\frac{6mT}{c^2\pi\hbar^2}\mathrm{Li}_{\frac{1}{2}} (-\mathrm{e}^{{A}/{T}})^2\right]  \right. \nonumber\\
&& \left. + \,\, \mathrm{Li}_{\frac{3}{2}} (-\mathrm{e}^{{A}/{T}})\left[1- \frac{12mT}{c^2\pi\hbar^2}\mathrm{Li}_{\frac{1}{2}} (-\mathrm{e}^{{A}/{T}})^2 \right]\right\}.\label{mu-d}
\end{eqnarray}
\end{widetext}
Finally, from the expressions (\ref{pressure-d}) and (\ref{mu-d}), we obtain the local pair correlations in the form
\begin{equation}
g^{(2)}(0) \approx -\frac{\gamma t^{\frac{3}{2}}}{\sqrt{\pi}}\mathrm{Li}_{\frac{3}{2}} (-\mathrm{e}^{{\tilde{A}}/{t}})
\left[1-\frac{11t}{\pi}\mathrm{Li}_{\frac{1}{2}} (-\mathrm{e}^{{\tilde{A}}/{t}}) \right] \label{g2-all}
\end{equation}
where $\tilde{A}$ is given by (\ref{EoS-E-A}).

We note that the local pair correlations satisfy a simple  relation
\begin{equation}
g^{(2)}(0) = \frac{2 p}{n\varepsilon}+O\left(\frac{1}{\gamma^4}\right),\label{g2-re}
\end{equation}
where $t$ is small. It turns out that this simple looking result not only holds for the quantum critical regime,
but also for the TLL phase and the ground state at $T=0$.
In this sense it is a universal relation.
The low temperature  local correlation for the TLL phase follows from (\ref{g2-re}) as
\begin{equation}
g^{(2)}(0)
= \frac{4\pi^2}{3\gamma^2}\left[1-\frac{6}{\gamma}+\frac{T^2}{4\pi^2\left( \frac{\hbar^2 }{2m}n^2 \right)^2}\right]+O\left(\frac{1}{\gamma^4}\right), \label{g2-TLL}
\end{equation}
which coincides with the result given in \cite{Kheruntsyan,Cazalilla2,GBT,Mussardo}.

Fig.~\ref{fig:g2-all} shows a plot of the local pair correlations obtained from (\ref{g2-all}) in comparison with the numerical result.
The crossover temperatures, which separate the quantum critical regime from the TLL,
are determined by the breakdown of the TLL with the local correlation (\ref{g2-TLL}).
The crossover line  is consistent with the phase diagram Fig.~\ref{fig:scaling}.
The TLL phase persists below the crossover temperatures, where both results (\ref{g2-re}) and (\ref{g2-TLL}) coincide, see  Fig.~\ref{fig:g2-all}.
This represents a smooth crossover from the relativistic TLL regime to the regime governed by a nonrelativistic dispersion relation.
The local correlations increase as the chemical potential becomes large and positive because of the decrease of the interaction.
The  temperature  enhances the local pair correlations, see Fig.~\ref{fig:g2TLL}.
We see clearly that the local correlations  (\ref{g2-re}) cover the quantum critical region,  the TLL phase and ground state at $T=0$.

For nonzero temperature, the vacuum can be taken
  as a semi-classical gas regime,  where the  particle density  $n \sim \frac{1}{\lambda }e^{-|\mu|/T}$ with thermal
  wavelength $\lambda^{-1}=\sqrt{m k_BT/2\pi\hbar^2}$, which  is much smaller than the mean distance between two particles.
  The local pair correlation tends to zero.
We see that the relation  (\ref{g2-re}) between the local correlations and pressure  holds in the physical regime as long as the
temperature is below the degenerate temperature and $T\ll \frac{\hbar^2}{2m}c^2$.
Near the critical point $\mu_c=0$, the density is very low, therefore the interaction  is strong and sits in the Tonks-Giraradau regime.
 At high temperatures, this crossover disappears due to suppression of the quantum fluctuations.
 At high temperatures, the chemical potential become more negative. We consider this limit in the next section.

\section{High temperature expansion}

The recent measurements \cite{Armijo} on thermal fluctuations in the 1D Bose gas  were carried out in the weak coupling and  high temperature regimes.
The variance of atom number $\langle \delta N^2\rangle$  in a volume $\Delta$ can be evaluated as \cite{Armijo}
\begin{equation}
\langle \delta N^2\rangle=\Delta k_B T \left(\frac{ \partial n(\mu, T) }{\partial \mu}  \right)_T
\end{equation}
where the local density $n(\mu, T)$ can be determined from the equation of state in the thermodynamic limit.  
In particular, Armijo \cite{Armijo2} has  recently  observed the quantum phonon fluctuations in the 1D Bose gas using in situ absorption imaging and statistical analysis of the density profiles. 
This opens up further study  of quantum vacuum fluctuations in a finite Lieb-Liniger Bose gas.

Here we first consider the equation of state of  this system at   high temperatures and in strong coupling regimes.
At high temperatures, the TBA equation (\ref{TBA}) can be expanded in an appropriate form.
For convenience, we define the function $\eta(k)=\mathrm{e}^{\epsilon(k)/T}$ and the inverse temperature parameter $\alpha=\hbar^2c^2/(2mT)$.
The TBA equation (\ref{TBA}) can then be written
\begin{equation}
\eta^{-1}(k)=x_k \, \mathrm{e}^{\int_{-\infty}^{\infty} dq \, a_2(k-q)\ln\left(1+\eta ^{-1} (q) \right)},\label{high-T}
\end{equation}
where $x_k=\mathrm{e}^{(\mu-\epsilon_0(k))/T}\ll 1$ at  high temperatures.
We carry out an expansion of this equation in powers of $Z=\mathrm{e}^{\mu/T}\ll 1$ as $T\to \infty$.
Collecting the first few terms, we have
\begin{eqnarray}
\eta^{-1}(k)\approx x_k\left(1+B_1Z+B_2Z^2+B_3Z^3 \right) +O(Z^5) \label{eta}
\end{eqnarray}
where the coefficients $B_i$,  for finite $\alpha$,  are given by
\begin{widetext}
\begin{eqnarray}
B_1&=&\frac{c}{(c^2+k^2)\sqrt{\alpha \pi} },\nonumber\\
B_2&=& \frac{1}{(c^2+k^2)\alpha \pi }+ \frac{c^2}{2(c^2+k^2)^2\alpha \pi } - \frac{c}{2\sqrt{2}(c^2+k^2)\sqrt{\alpha \pi} },\nonumber\\
B_3&=& \frac{c}{(c^2+k^2)^2(\alpha \pi)^{\frac{3}{2}} }+ \frac{c^3}{6(c^2+k^2)^3(\alpha \pi)^{\frac{3}{2}} }
- \frac{1}{\sqrt{2}(c^2+k^2)\alpha \pi }-\frac{c^2}{2\sqrt{2}(c^2+k^2)^2\alpha \pi }.
\end{eqnarray}

Using the expansion (\ref{eta}) gives the pressure (\ref{Pressure}) in the asymptotic form
\begin{eqnarray}
p= \sqrt{\frac{m}{2\pi\hbar^2}}T^{-\frac{3}{2}}Z\left[1+f_1Z+f_2Z^2+f_3Z^3\right]+O(Z^5),\label{highT}
\end{eqnarray}
with
\begin{eqnarray}
f_1&=& \mathrm{e}^{\alpha} \left( 1-\mathrm{Erf}(\sqrt{\alpha})\right)-\frac{1}{2\sqrt{2}},\nonumber\\
f_2&=& \mathrm{e}^{\alpha}\left( 1-\mathrm{Erf}(\sqrt{\alpha})\right)\left(\frac{5}{4\sqrt{\pi\alpha}}-\frac{\sqrt{\alpha} }{2\sqrt{\pi}}-\frac{1}{2\sqrt{2}} \right)
- \mathrm{e}^{2\alpha}\left( 1-\mathrm{Erf}(\sqrt{2\alpha})\right)+\frac{1}{2\pi}+\frac{1}{3\sqrt{3}},\nonumber\\
f_3&=&\frac{9}{8\pi^{\frac{3}{2}}\sqrt{\alpha} }-\frac{\sqrt{\alpha} }{12\pi^{\frac{3}{2}}}-\frac{1}{2\sqrt{2}\pi }-\frac{\sqrt{2}}{\pi}\nonumber\\
&&+\, \mathrm{e}^{\alpha} \left( 1-\mathrm{Erf}(\sqrt{\alpha})\right)\left(\frac{1}{2\pi^{\frac{3}{2}}\alpha}-\frac{11-\alpha }{12\pi}
 +\frac{1}{16\pi\alpha }-\frac{5}{4\sqrt{2\pi\alpha}}+\frac{\sqrt{\alpha}}{2\sqrt{2\pi}}+\frac{1}{3\sqrt{3}}\right)\nonumber\\
&&+\,  \mathrm{e}^{2\alpha}\left( 1-\mathrm{Erf}(\sqrt{2\alpha})\right)\left( \frac{1}{2\sqrt{2}}-\frac{3}{2\sqrt{\pi\alpha }}-\frac{2\sqrt{\alpha} }{\sqrt{\pi} }\right)
+ \mathrm{e}^{3\alpha} \left( 1-\mathrm{Erf}(\sqrt{3\alpha})\right) ,
\end{eqnarray}
where $\mathrm{Erf}(x)$ is the standard error function.
\end{widetext}

The TBA results  reduce to several limiting cases.
The pressure $p_c$ of the classical Boltzmann gas
\begin{equation}
p_c= \sqrt{\frac{m}{2\pi\hbar^2}} \, T^{-{3}/{2}} \, \mathrm{e}^{{\mu}/{T}}
\end{equation}
follows in the limit $T\to \infty$. The pressure of the ideal Fermi gas
\begin{equation}
p=-\sqrt{\frac{m}{2\pi\hbar^2}} \, T^{{3}/{2}} \, \mathrm{Li}_{\frac{3}{2}} (-\mathrm{e}^{{\mu}/{T}})
\end{equation}
is obtained in the limit $c \to \infty$. Similarly the pressure of the ideal Bose gas
\begin{equation}
p=\sqrt{\frac{m}{2\pi\hbar^2}} \, T^{{3}/{2}} \, \mathrm{Li}_{\frac{3}{2}} (\mathrm{e}^{{\mu}/{T}})\label{pressure-boson}
\end{equation}
follows in the limit $c\to 0$.

\begin{figure}[tbp]
\includegraphics[width=1.1\linewidth]{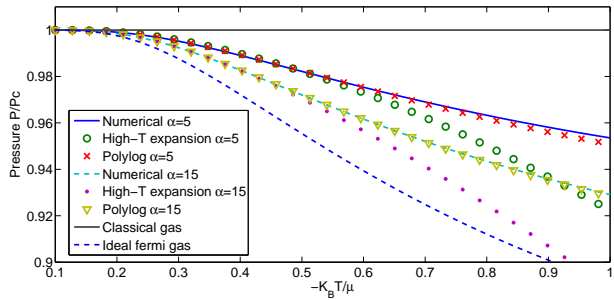}
  \caption{The ratio of the pressure $p/p_c$ vs $k_BT/\mu$ at high temperatures for $\alpha=5$ and $15$.
  Here $p_c$ is the pressure of the classical Boltzmann gas.
High accuracy of the pressure  (\ref{pressure-polylog})  is seen for $-\infty < k_BT/\mu <1$. 
At high temperatures both  pressure   (\ref{pressure-polylog}) obtained from the polylog formalism  and  the one (\ref{highT}) 
obtained from high temperature  expansion  are accurate by comparing with the  numerical result  from the TBA equation (\ref{TBA}). 
The polylog formalism is also valid at low temperatures as long as $\alpha \gg 1$.  }
\label{fig:highT}
\end{figure}

In the high temperature limit, the chemical potential tends to negative infinity.
In this region, far away from criticality, the pressure (\ref{pressure-polylog}) given in terms of the polylog function and  the
high temperature  expansion result (\ref{highT}) are highly accurate for $\alpha >5$,  see Fig.~\ref{fig:highT}.
The polylog function result (\ref{pressure-polylog}) gives a better fit with the numerical result obtained from the TBA (\ref{TBA})
even for large values of $k_BT/\mu$.  At high temperatures, although the thermal fluctuations dominate, the quantum statistics are
still microscopically significant.  The quantum statistical effect  is evidenced from changing the value of $\alpha$,
which can be controlled in current experiments \cite{Armijo,Armijo2}.
In the Tonks-Girardeau limit $\alpha \to \infty$ the gas approaches free fermions.
For small values of $\alpha$, i.e., in the week coupling region,  we see the pressure  (\ref{highT}) obtained from the high temperature  expansion
is consistent with the numerical result at high temperatures, see Fig.~\ref{fig:highT}.
But as expected, the result (\ref{highT})  is no longer accurate  for $\alpha \ll 1$.

\begin{figure}[tbp]
\includegraphics[width=1.1\linewidth]{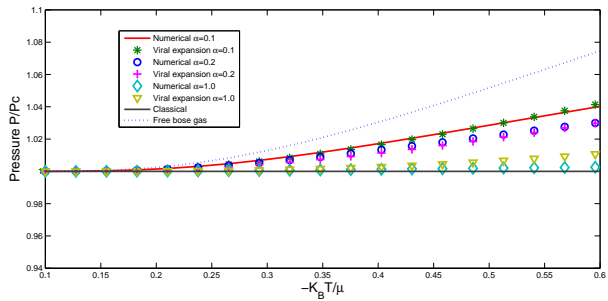}
  \caption{Comparison between different values for the pressure $p/p_c$ vs $k_BT/\mu$ at high temperatures for $\alpha=0.1,\,0.2,\,1.0$.
  For small values of $\alpha $,  the pressure  (\ref{Viral}) obtained from the virial  expansion  is  in good agreement  with  the
  result obtained  from the numerical solution of the TBA equation (\ref{TBA}). }
\label{fig:highT-L}
\end{figure}

In the weak coupling limit and at high temperatures it is more practical to consider a virial expansion with the TBA equation (\ref{TBA}), i.e.,
\begin{equation}
p\sqrt{\frac{2\pi\hbar^2}{m}}T^{\frac{3}{2}}={\rm Li}_{\frac{3}{2}}\left(Z\right)+\sqrt{2} \, p_2 Z^2 +O(Z^3) \label{Viral}
\end{equation}
with $Z=\exp(\mu/T)$.
The first term in the rhs of this equation comes from free bosons and
\begin{equation}
p_2=-\frac{1}{2}-\frac{\exp(\alpha^2/2)}{2}\left[-1+{\rm Erf}\left(\sqrt{\alpha/2} \right) \right]
\end{equation}
contributes to the second virial coefficient.
From the comparison shown in Fig.~\ref{fig:highT-L}, we see clearly that the pressure obtained from the virial expansion
gives a high precision equation of state for $\alpha < 1$.  This result is valid for the experimental setting with $\alpha <1$ \cite{Armijo,Armijo2}.

\section{Conclusion and discussion}

In conclusion, we have studied finite temperature local pair correlations of the Lieb-Liniger Bose gas at quantum criticality.
A simple relation (\ref{g2-re}) between the local correlations and the pressure has been obtained
in the framework of the Yang-Yang  thermodynamic equations.
This relation holds for both the quantum critical regime and the TLL phase at quantum criticality.
It provides a simple way to probe  finite temperature local pair correlations for  the TLL over the whole relativistic  dispersion regime
and to test  quantum criticality with the local correlations  beyond the TLL phase.
In the quantum critical regime, the thermal fluctuations strongly couple to the quantum fluctuations with universal free fermion $z=2$ quantum criticality.
The local pair correlations provide insight into the microscopic quantum statistical effects at quantum criticality.
We also derived the thermodynamic equations (\ref{highT}) and (\ref{Viral}) of the Lieb-Liniger gas at high temperatures using a high temperature expansion of the
Yang-Yang  thermodynamic equations. The effect of quantum statistics  is microscopically significant even in the thermal fluctuation
dominated high temperature regime. In particular, the equation of state given by  (\ref{Viral}) is highly accurate for  the weak coupling and high temperature regimes.

Our analytical prediction (\ref{g2-re}) for local pair correlations can be tested using current experimental techniques for preparing and detecting 1D gases.
An ensemble of parallel 1D Bose gases can be prepared \cite{Kinoshita,Exp2} by loading ultracold Bose atoms into a two-dimensional (2D) optical lattice.
The 2D optical lattices, an ensemble of parallel 1D tubes, can be formed by superimposing two standing-wave lasers on the crossed dipole trap.
The depth of the 2D lattice must be sufficiently large to make the quantum tunnelling among these 1D tubes negligible.
That is, atoms in these 1D tubes are almost all in the lowest transverse vibrational state.
The photoassociation techniques for measuring local pair correlations in zero-temperature 1D Bose gases \cite{Exp2}
could also be used to explore local pair correlations in non-zero temperature systems,
such as the Lieb-Liniger Bose gas at quantum criticality studied in this paper.
In particular, the universal relation between the local pair correlation and the pressure at quantum criticality could be explored following the
experimental scheme for measuring the homogeneous contact of a unitary Fermi gas \cite{Sagi}.

\section*{Acknowledgements}
This work has been supported by the Australian Research Council and by the National Basic Research Program of China 
under Grant No. 2012CB821305 and No. 2012CB922101,
the National Natural Science Foundation of China under Grant No. 11075223, 
the NCETPC under Grant No. NCET-10-0850.  
X.-W.G. thanks Sun Yat-Sen University  for their hospitality during his visits.  
M.-S.W. thanks the Australian National University for kind hospitality.

\end{document}